\documentclass[prl,10pt,twocolumn]{revtex4}

\usepackage{amsfonts,amsmath}
\usepackage{float}
\usepackage{graphicx}   

\newcommand{\lac}{\textit{lac}}
\renewcommand{\subsection}[1]{\noindent\textbf{#1. }}

\begin{document}

\title{What makes the \textit{lac}-pathway switch: identifying the fluctuations that trigger phenotype switching in gene regulatory systems}

\author{Prasanna M. Bhogale$^{1\dag}$, 
Robin A. Sorg$^{2\dag}$,
Jan-Willem Veening$^{2*}$,
Johannes Berg$^{1*}$
}

\address{$^1$University of Cologne, Institute for Theoretical Physics,\\Z\"{u}lpicher Stra{\ss}e 77, 50937 K\"{o}ln, Germany\\
$^2$Molecular Genetics Group, Groningen Biomolecular Sciences and Biotechnology Institute, Centre for Synthetic Biology, University of Groningen, Nijenborgh 7, 9747 AG, Groningen, The Netherlands\\
$\dag$ these authors contributed equally\\
$^*$ correspondence to \texttt{J.W.Veening@rug.nl} and \texttt{berg@thp.uni-koeln.de}.
}

\begin{abstract}
Multistable gene regulatory systems sustain different levels of gene expression under identical external conditions. Such multistability is used
to encode phenotypic states in processes including nutrient uptake and persistence in bacteria, fate selection in viral infection,  cell cycle control, and development. Stochastic switching between different phenotypes can occur as the result of random fluctuations in molecular copy numbers of mRNA and proteins arising in transcription, translation, transport, and binding. However, which
component of a pathway triggers such a transition is generally not known. 
By linking single-cell experiments on the lactose-uptake pathway in \textit{E. coli} to molecular simulations, we devise a
general method to pinpoint the particular fluctuation driving phenotype switching and apply this method to the transition between the uninduced and induced states of the \lac~genes. 
We find that the transition to the induced state is not caused only by the single event of \lac-repressor unbinding, but depends crucially on the time period over which the repressor remains unbound from the \lac-operon. We confirm this notion in strains with a high expression level of the repressor (leading to shorter periods over which the \lac-operon remains unbound), which show a reduced switching rate. 
Our techniques apply to multi-stable gene regulatory systems in general and allow to identify the molecular mechanisms behind stochastic transitions in gene regulatory circuits. 
\end{abstract}

\maketitle

\section*{Introduction}

Multistable gene regulatory systems use specific mechanisms like feedback to stabilize expression patterns defining different phenotypic states~\cite{multavo,pmid23010999, pmid23781105, pmid22726437, pmid18388284, Isaacs2003, pmid16541134, pmid23372015}. However, no natural system is strictly multistable since it will not persist in any one of its states indefinitely. Instead, the lifetimes of stable states are finite, with random fluctuations causing transitions between different states. In gene regulatory systems the copy numbers of mRNA molecules, proteins, and ligands fluctuate over time due to the random timing of transcription, translation, transport, and binding~\cite{McAdamsArkin1996,EloSwain2002,pmid15687275,blakeetal2003,metozvo,pmid17517669,pmid14749823, HIVweinberger}. 
These fluctuations can trigger a switch from one phenotype to another~\cite{pmid17569828, arabinonsw, xieonmod, Cagatayetal2009,ZongGolding2010}. 
However, when all components of a multistable system fluctuate, what are the fluctuations causing the transition? In other words, what 
are the rate-limiting fluctuations?

\begin{figure}[tbh]
\begin{center}
\includegraphics*[width=0.5\textwidth]{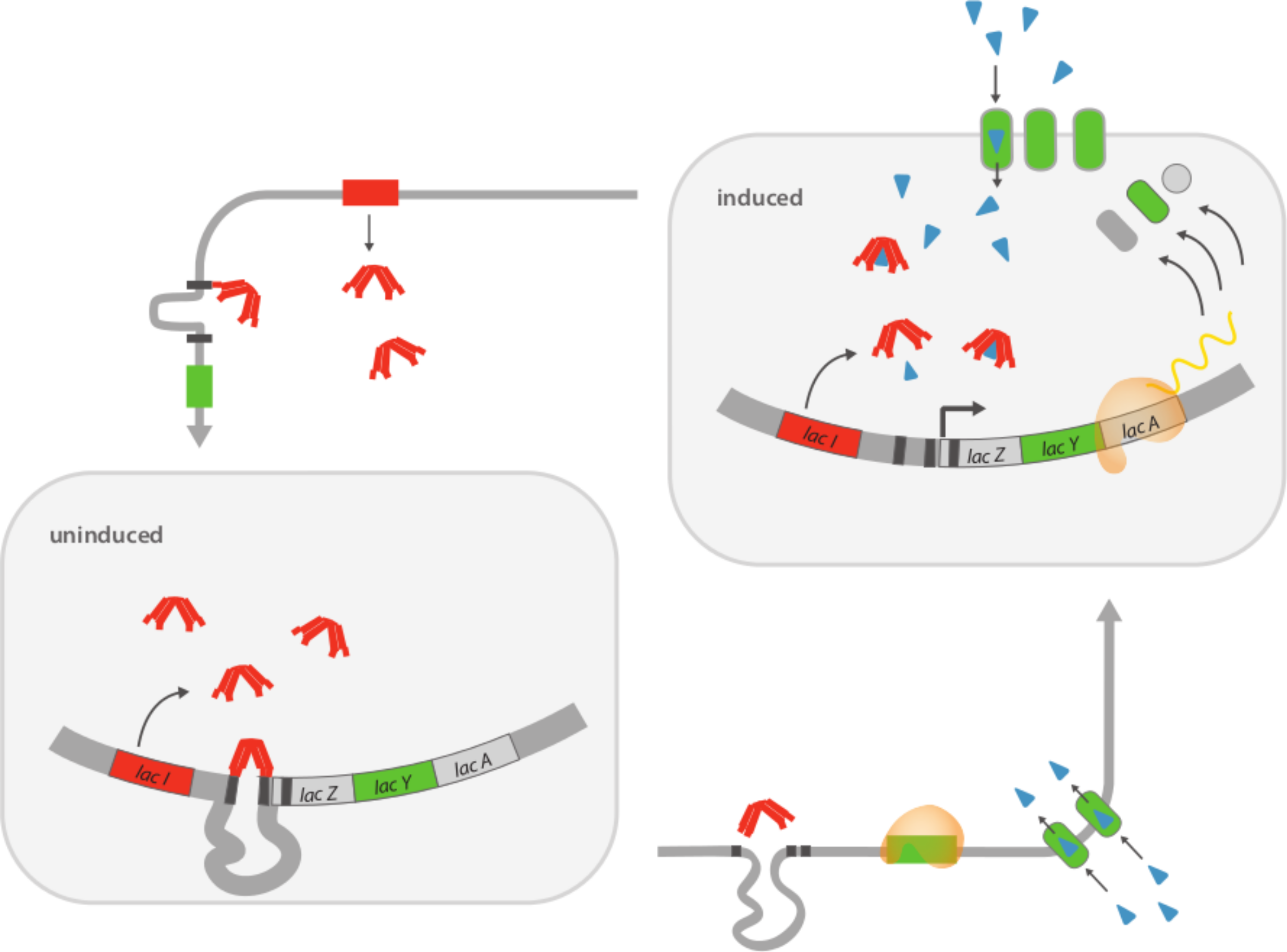}
\caption{{\bf Feedback in the lactose-uptake pathway.} Three genes under common regulatory control constitute the lactose-uptake pathway:
\textit{lacZ} encodes the enzyme to break down lactose, \textit{lacY} encodes a permease (importer of lactose, shown in green), which actively imports lactose from the environment  (here: the non-metabolizable lactose-analogue TMG, blue wedge), and \textit{lacA} encodes a transacetylase. Bistability is achieved by a positive feedback-loop; the imported TMG acts as an inducer of the \lac~genes by increasing the unbinding rate of LacI tetramers (red) from the so-called operator binding sites in the \lac-regulatory region. LacI tetramers inhibit the expression of the \lac~genes. As a result, the induced state shown on the right, with the \lac~genes expressed, is stable if the permease imports enough inducers to deactivate the repressors. 
The uninduced state shown on the left can also be stable, since in the absence of~\lac-expression, the repressors prevent expression of the \lac~genes by binding to 
the regulatory region and forming a DNA loop. However, this regulatory circuit is built from stochastic components and can be interrupted by 
random fluctuations of the number of mRNA, proteins, inducer, repressors, or the binding state of repressors to DNA. These fluctuations lead to transitions 
between the induced and uninduced states.}
\label{fig:lacsystem}
\end{center}
\end{figure}

Such rate-limiting fluctuations are difficult to identify experimentally, because it is hard to monitor and control variation in molecular copy numbers inside a cell. 
In an ideal scenario, one would take a multistable system and reduce the amplitude of fluctuations of each of its components in turn. 
If reducing the fluctuations of a particular component affects the switching rate between different states, then one can consider fluctuations in that component rate-limiting to the transition. 
This strategy has been implemented experimentally by Maamar \textit{et al.}~\cite{pmid17569828} for a single component of a bistable signaling pathway in \textit{Bacillus subtilis}. 
The \textit{comK}-pathway enables \textit{B. subtilis} to take up new genetic material, which may offer fitness advantages~\cite{pmid19189946}.
Maamar \textit{et al.} increased the transcription rate of \textit{comK}  and simultaneously decreased its translation rate. Average protein levels were left unaffected, but fluctuations around this mean due to the random timing of mRNA production were reduced. Maamar \textit{et al.} observed a decrease in the switching rate between the states with low levels of ComK and high levels of ComK, showing that mRNA-fluctuations affect the switching rate. But are these the only rate-limiting fluctuations in the system? Repeating this procedure for all components in a pathway is cumbersome for small pathways and infeasible for larger pathways. Moreover, transitions could be driven by fluctuations in ligand numbers, protein conformations or binding state, which are even harder to control in experiments.

If rate-limiting fluctuations are hard to identify experimentally, they cannot be identified purely on the basis of regulatory network models and computer simulations either. Any model describes a restricted number of molecular species and replaces the rest with effective reaction rates. The formulation of a model thus already constitutes an \textit{a priori} assumption on the relevant constituents. Being rare events, transitions between stable states are strongly influenced by the molecular details. Two models can thus exhibit the same bistable expression patterns as a function of external parameters, e.g. hysteresis plots~\cite{multavo}, yet differ markedly in the mechanisms and rates of switching between states. 
Indeed the rate-limiting fluctuation is thought to differ drastically across pathways: mRNA fluctuations in the \textit{comK}-pathway ~\cite{pmid17569828}, fluctuations in initial pump numbers in the arabinose-uptake pathway~\cite{arabinonsw}, and gene activity bursts in the $\lambda$-phage lysogeny~\cite{ZongGolding2010}. 
Hence, transitions between stable states can act as a sensitive probe into molecular details of a pathway.

The best-studied example of a multistable regulatory system is the lactose-uptake pathway in~\textit{Escherichia coli}, exhibiting bistable expression of the genes~\textit{lacZ, lacY, lacA}. 
Due to a positive feedback loop, both the induced state (high expression levels of the \lac-genes) and the uninduced state
 (low expression levels of the \lac-genes) can be stable, see Fig. \ref{fig:lacsystem}. 
The constituents and function of this pathway have been known for half a century~\cite{novic, mullerhillbook}, 
and most kinetic rates have been measured~\cite{lacoprep,sugimp,dunwaynum,repdnabin,riggs1970,friedman1977}, and the parameter space over which the system is bistable has been explored~\cite{multavo}. Despite a large body of experimental~\cite{
multavo,
metozvo,
epiinrobert,
xieonmod}
and computational studies~\cite{StamatakisZygourakis2011, Robertsetal2010,  pmid21414326}, there is no consensus on the transition mechanism. Experimentally, a key observation was made by Choi~\textit{et al.}~\cite{xieonmod}, who find that the \lac-repressor tetramer is released just before the transition to the induced state. But is this single-molecule event alone driving the transition, or are there other fluctuations involved? A study by Robert \textit{et al.} \cite{epiinrobert} shows that cellular concentrations of the repressor affect the switching behavior of the \lac-pathway, suggesting that rebinding of the repressor to the operator might play a role.  

In this paper, we use single cell analysis to determine switching rates from the uninduced to the induced state of the \lac-system. We set up a detailed mechanistic model of the \lac-system
and propose a simple scheme to reduce fluctuations in the constituents of the \lac-pathway~\textit{in silico} and identify the fluctuations driving the transitions between states. Our method to pinpoint rate-limiting fluctuations is general and can be applied to any system whose kinetic rates are sufficiently well known. 

\section*{Materials and methods}

\subsection{{Determining phenotypic switching rates}}
\begin{figure*}[tbh]
\begin{center}
\includegraphics*[width=0.96\linewidth]{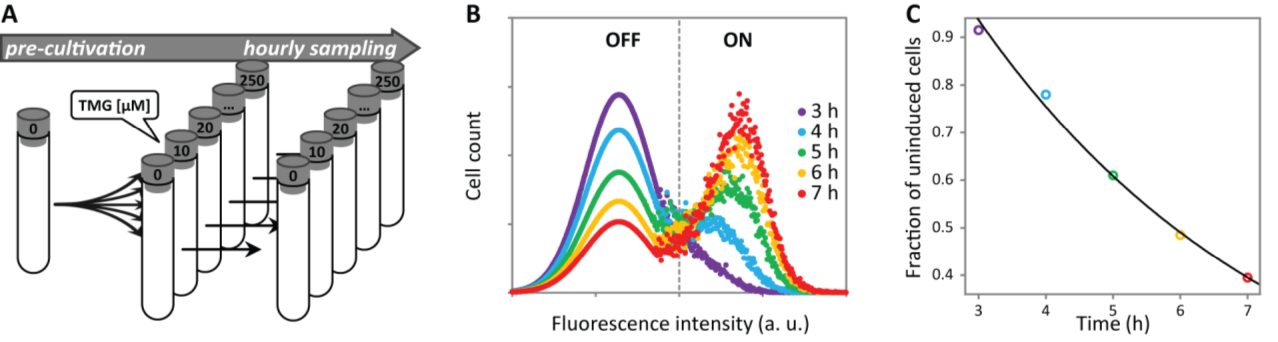}
\caption{\textbf{Single-cell analysis and the rate of phenotype switching.} 
\textbf{A:} (Schematic) We take hourly samples from populations of \textit{E. coli}  cells growing at different concentrations of the external inducer. Through dilution at regular intervals we keep the populations under constant conditions, see Materials and Methods and SI.  A fluorescent reporter indicates the expression levels of the~\lac~genes in individual cells. 
\textbf{B:} Fluorescence data taken at $30\mu$M of TMG (smoothed with a moving average filter for visual clarity) showing the bimodal distribution of reporter expression with high and low levels corresponding to cells in uninduced (OFF) and induced (ON) states. Initially, all cells are in the uninduced state. With time, the fraction of cells in the induced state increases and the fraction of cells in the uninduced state decreases (shown here: purple 3h, blue 4h, green 5h, yellow 6h, red 7h, other time points not shown; for data processing see SI).
For images of single cells and their fluorescence over time, see Fig~\ref{fig:plasmid}.
We observe no significant difference between growth rates of induced and uninduced cells, see SI. 
\textbf{C:} The fraction of cells in the uninduced state decays approximately exponentially with time. Fitting an exponential function (black line) to the data points gives the switching rate to the induced state. For this particular concentration of TMG the switching rate is $3.9\times10^{-3} \pm 3.3\times 10^{-4}/min$. }
\label{fig:single_cell} 
\end{center}
\end{figure*}
To assess expression of the \lac~genes at the single-cell level we used flow cytometry on a population of \textit{E. coli} strain CH458, which contains a \textit{gfp-cat} cassette inserted downstream of the \lac-operon ~\cite{errgordon} (see SI section \textit{1} for details). The switching rate from the uninduced state to the induced state is the number of cells per unit time which switch from low numbers of \lac-proteins to a state with high number of \lac-proteins. To determine the rate of switching to the induced state, we start with a population of uninduced cells and fit the fraction of cells in the uninduced state at subsequent times to an exponential decay. Examples are shown in Figure {\ref{fig:single_cell}} and Figure S3. Switching rates are determined at different external inducer concentrations, resulting in a rate curve of the switching rate against inducer concentration (see Figure \ref{fig:transitionrate}). The reverse transition from the induced to the uninduced state is slow by comparison and, on the timescales of our experiment, few cells switch back from the induced to the uninduced state.

\subsection{Experimental conditions}
We used the non-metabolizable thio-methylgalactoside (TMG) as an inducer. 
M9 minimal salts supplemented with thiamine, MgSO$_4$, CaCl$_2$, and casamino acids were chosen as growth medium for the CH458 strain used in our experiments. We used succinate instead of glucose as carbon source to reduce catabolite repression yet maintain sufficient growth. Chloramphenicol (10 $\mu$ g/ml) was added to reduce the risk of contamination during sampling. Plasmid pREP4 (Qiagen) was transformed to strain CH458 using CaCl$_2$ transformation and selecting for kanamycin resistance, resulting in strain CH458+pREP4. 

For rapid and robust optical density (OD) determination, cultures of 2ml were grown in 5ml tubes fitting directly into the spectrophotometer. Overnight cultures were grown at different TMG concentrations for 16 hours (approximately 12 generations) at 37$^{\circ}$C while shaking. In the morning, uninduced cells from 0 $\mu$M TMG and induced cells from 250 $\mu$M TMG were diluted to OD 0.04 and grown in the particular TMG concentration for four additional hours. Cultures were re-diluted every hour to ensure constant exponential growth within a narrow OD range between 0.03 and 0.08. After this adaptation phase, cells were spun down and washed for the change of medium to the desired final TMG concentration. In the following eight hours, samples were taken every hour for measurement and cultures were rediluted as described above to maintain steady state. Fluorescence of individual cells was determined by flow cytometry of $10^5$ cells at a time (see Supplementary Figure S1). The fluorescence intensity was measured with a BD FACS Canto Flow Cytometer of BD Bioscience at medium flow with a forward scatter (FSC) of 200V and side scatter (SSC) of 400V. A threshold of 500V was set to exclude recording of particles that are smaller than normal \textit{E. coli} cells. The fluorescence detector FL-1 was set to 800V for a maximal separation of induced versus uninduced cells. We found it crucial to maintain constant growth conditions by dilution to keep intra and extra-cellular parameters stable. 

Time-lapse microscopy was carried out by spotting single cells on a M9 polyacrylamide (10\%) slide containing TMG inside a Gene Frame (Thermo Fisher Scientific). A Nikon Ti-E microscope equipped with a CoolsnapHQ2 camera and an Intensilight light source was used in an Okolab climate incubator at 37$^{\circ}$C. Images with cells expressing the green fluorescent protein (GFP) were taken with the following protocol and filter set: 200ms exposure time for phase contrast and 0.5s exposure for fluorescence at 450-490 nm excitation via a dichroic mirror of 495nm and an emission filter at 500-550 nm. Pictures were taken every 30 minutes. 

\subsection{{Identification of rate limiting fluctuations via smoothing \textit{in silico}}}
{We set up a detailed mechanistic model of the lactose-uptake pathway which captures the switching behavior observed experimentally.} This model describes the expression of mRNA and protein, both of LacY (lactose/TMG importer) and the repressor (LacI tetramer), the uptake of external inducers into a cell, repressor binding to DNA, DNA looping, passive diffusion of inducers into the cell, and cell division. 
All rate constants were taken from the literature, with the exception of a single parameter describing the rate of inducer import. For this parameter, literature values vary widely
and may depend strongly on the particular strain used. This parameter is the sole free parameter of our model and is determined by fitting the experimentally determined switching rates to those observed in the mechanistic model. For details of the modeling and the literature sources of rate constants, see SI section \textit{2}. 

\begin{figure}[b!]
\begin{center}
\includegraphics*[width = 0.5\textwidth]{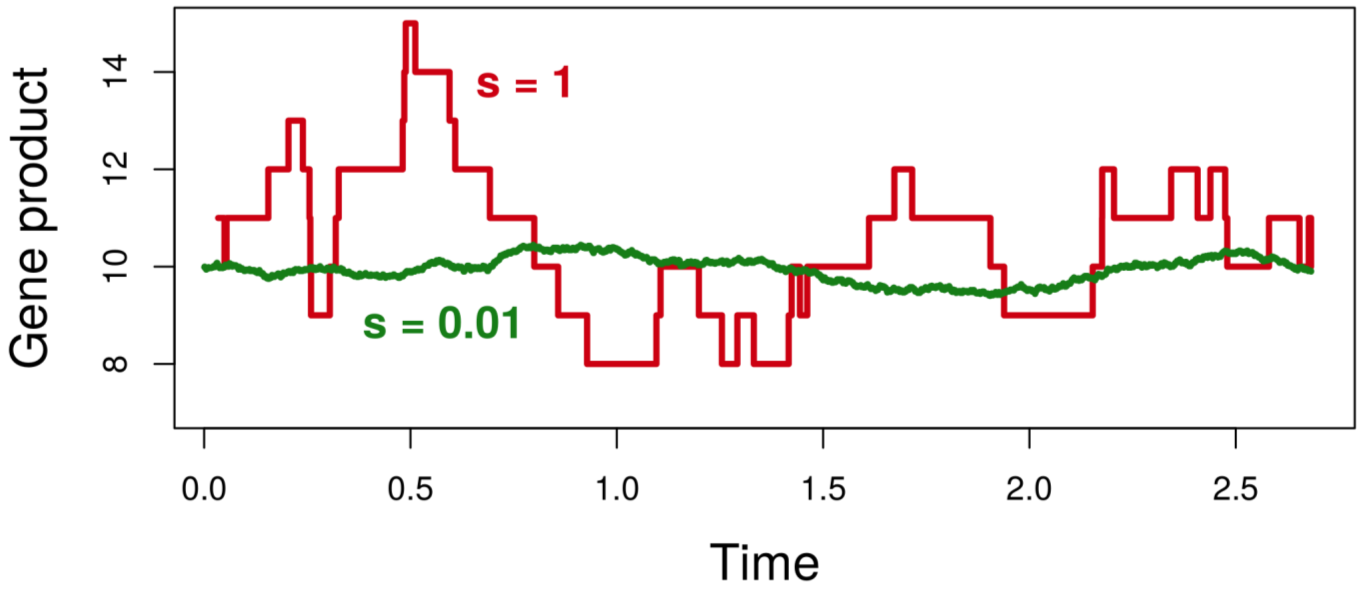}
\caption{{\bf \textit{In silico}-smoothing of fluctuations in gene regulatory systems.} As an example we show the production and degradation of some molecule at constant rates (red line). In order to reduce fluctuations, 
 the step size (one molecule at a time in the original dynamics) is multiplied by a smoothing factor $s<1$, but the rate at which the smaller steps occur is increased by $1/s$. The resulting dynamics (green line $s=0.01$) has the same mean, but the amplitude of fluctuations is reduced by a factor proportional to $s$ relative to the 
original dynamics.}
\label{fig:smoothing}
\end{center} 
\end{figure}

{To assess the effect of fluctuations in specific components of the lac-pathway on the switching rate, we put forward a simple scheme to control fluctuation amplitudes \textit{in silico}.}
Take a particular component, e.g. \textit{lacY}-mRNA that is produced in units of $1$ molecule at some rate. (This rate changes over time due to repressor binding and unbinding to the \lac~regulatory region.) We now change the number of molecules produced in each transcription event by a smoothing factor $s<1$, and simultaneously divide the transcription and degradation rates by the same factor. For a smoothing factor $s=0.1$, mRNA molecules are produced in units of $1/10$ but at $10$ times the rate. 
This is impossible to do experimentally but feasible \textit{in silico}; downstream, the rate of protein production will now simply be proportional to a non-integer number of mRNA molecules, and analogously for other molecules and binding states. The effect of this procedure is illustrated in Fig.  {\ref{fig:smoothing}}, see also SI Section \textit{3}. The mean number of molecules is preserved, but the fluctuations about this mean are reduced by a factor of $s$. Deterministic dynamics (which can be described by ODEs) corresponds to $s=0$, while finite values of $s$ result in some degree of stochasticity. If smoothing fluctuations in a particular component affects the switching rate between phenotypic states, we conclude that these fluctuations are rate limiting to the particular transition. On the other hand, if smoothing fluctuations in a particular component is found not to affect the switching rate, its dynamics can be modeled by an ordinary (deterministic) differential equation. In this way, the minimal model describing a particular transition can be determined systematically. 

\section*{Results}

\subsection{{Mechanism of switching in the \lac-system}}
\begin{figure}[bt]
\begin{center}
\includegraphics*[width=.5\textwidth]{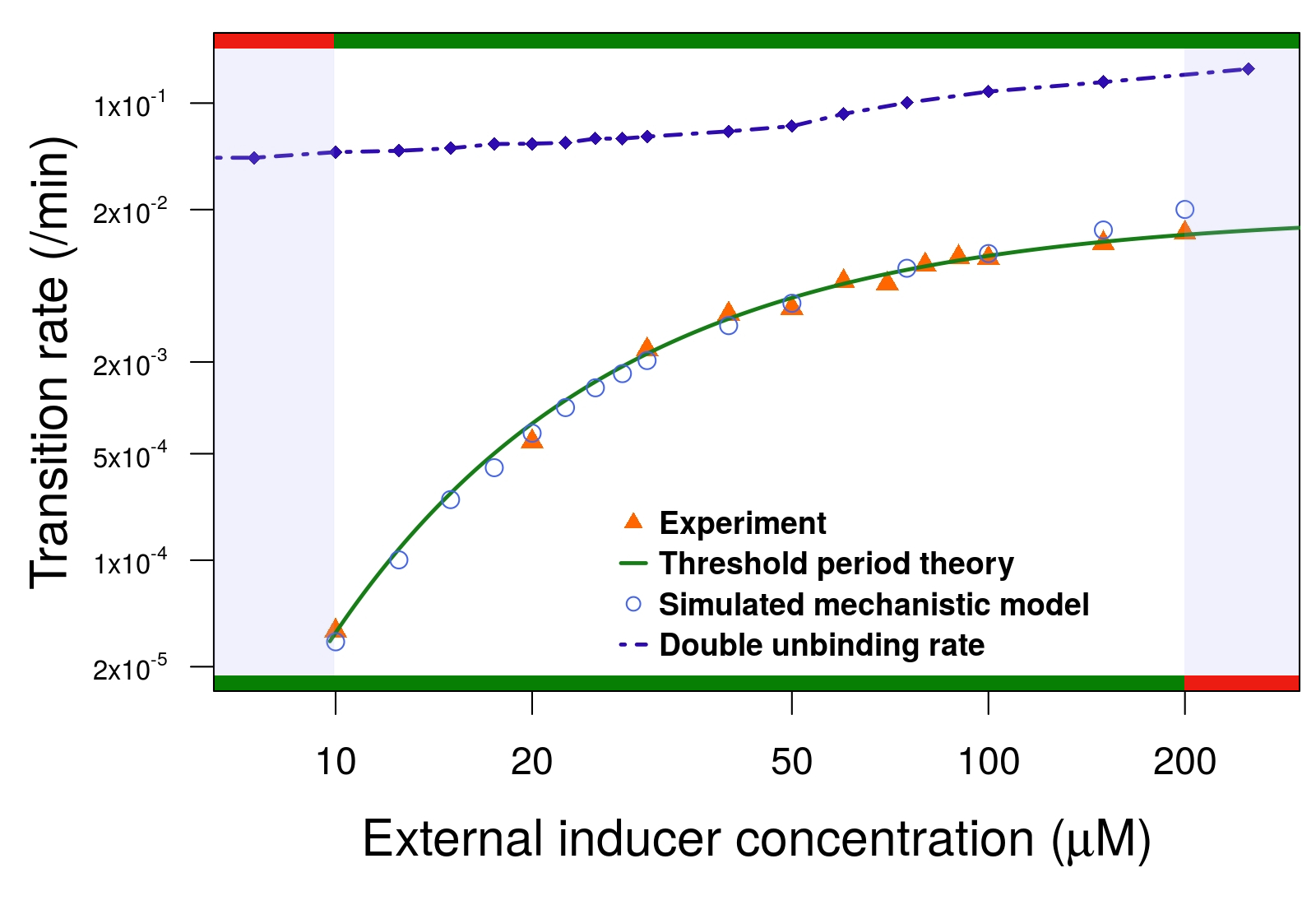}
\caption{{\bf The transition from uninduced to induced state of the \lac-system.} The unshaded region indicates the range of external inducer (TMG) concentration where switching rates from the uninduced to the induced state of the \lac-system could be determined experimentally. 
Orange triangles give the switching rates as determined in Fig. {\ref{fig:single_cell}} for different inducer concentrations. The experimentally determined switching rates agree closely with the switching rates observed in computer simulations of the mechanistic model (blue circles). 
The rate of unbinding of the repressor from both its binding sites (simulations, dashed blue line) which was established by Choi \textit{et al.}~\cite{xieonmod} as a necessary condition for a switch to occur, is up to $3$ orders of magnitude greater than the switching rate. 
The full green line gives our theoretical result \textit{(\ref{eq:rate_switchon})} for the switching rate, which takes into account the threshold time period between repressor unbinding and rebinding, see text.} 
\label{fig:transitionrate}
\end{center} 
\end{figure}

For external concentrations of the inducer TMG between $7.5\mu M$ and $200\mu M$ the \lac-system shows bistability with states of high and low expression levels of the \lac~genes {(see Fig. S{2})}. By measuring the relative numbers of induced and uninduced cells in the population over time we determine the switching rate per cell at a given external TMG concentration, see Figure. {\ref{fig:single_cell}}. 
Figure. {\ref{fig:transitionrate}} shows the switching rate from the uninduced to the induced state against the external TMG concentration (orange triangles). The excellent match of this switching rate curve with that found in numerical simulation of the mechanistic model (blue circles) suggests that our model contains the relevant components of the \lac-pathway. In order to pinpoint the rate-limiting fluctuations, we then reduce the fluctuations of each component of our model in turn as described above and in Fig. {\ref{fig:smoothing}}.

We find that the switching rate curve is unaffected by reducing fluctuations in all components except the operator state. Notably, fluctuations arising from the finite number of \textit{lacY} mRNA and protein, partitioning due to cell division, or the random timing of transcription and translation do not affect the switch to the induced state. 
Only the random timing of repressors binding to and unbinding from the \lac-operator affects the switching rate. Reducing these fluctuations, the switching rate decreases until no more transitions to the induced state are observed
on the timescale of our simulations. On the other hand, the stochastic dynamics of all other components can be replaced by a smooth deterministic dynamics without affecting the switching rate (see SI Section \textit{3}). Fluctuations in the operator state are thus the rate-limiting fluctuations for the transition to the induced state. 

In the \lac-system, LacI molecules can tetramerize and bind simultaneously to two different operator sites, forming a DNA loop that results in a very effective repression of transcription. A mechanism proposed by Choi \textit{et al.}~\cite{xieonmod} is that the repressor unbinds from both operator sites,  triggering a burst of mRNA production taking the \lac-pathway to the induced state. However, we find that full repressor unbinding takes place at a far higher rate than the transition to the induced state (see Fig.  {\ref{fig:transitionrate}} and SI Section \textit{4}). An additional mechanism must be involved. 

Once the repressor has released the \lac-operon, the same repressor molecule (or a different one) might quickly bind again. Alternatively, the operator might remain unbound for a time period $\tau$ long enough for sufficient numbers of LacY to be produced and for sufficient inducer molecules to be pumped into the cell to deactivate repressors and switch the cell to the induced state. In our simulations we find that, upon repressor unbinding, the transition to the induced state takes place only if the time period over which the operator site remains unbound exceeds some threshold. 
Accounting for this threshold period gives a simple but accurate theory of the transition to the induced state of the \lac-system, analogous to the theory developed by Walczak, Onuchic and Wolynes for a simple model of a self-activating gene~\cite{walczakwolynes}.

The effect of \lac~expression on the repressors of the \lac~genes is mediated by the inducing sugars imported by the importer LacY. The threshold period $\tau$ is then given by the time  required to express a number of importers sufficient to maintain a certain concentration of inducers in the cell. This critical number of importers is set by the requirement that repressors are deactivated by binding to the imported inducers. We find the threshold period $\tau$ depends on the external concentration of inducers $i$ via
\begin{equation}
{\tau = \beta \frac{i+i_0}{i} \ ,}\label{eq:an_tau}
\end{equation}
where $i_0$ is the half maximum of inducer import while $\beta$ is proportional to mRNA and protein dilution rates and inversely proportional to rates of translation, transcription and sugar import by proteins (see SI Section \textit{4}).

\begin{figure}[bt]
\begin{center}
\includegraphics*[width=.45\textwidth]{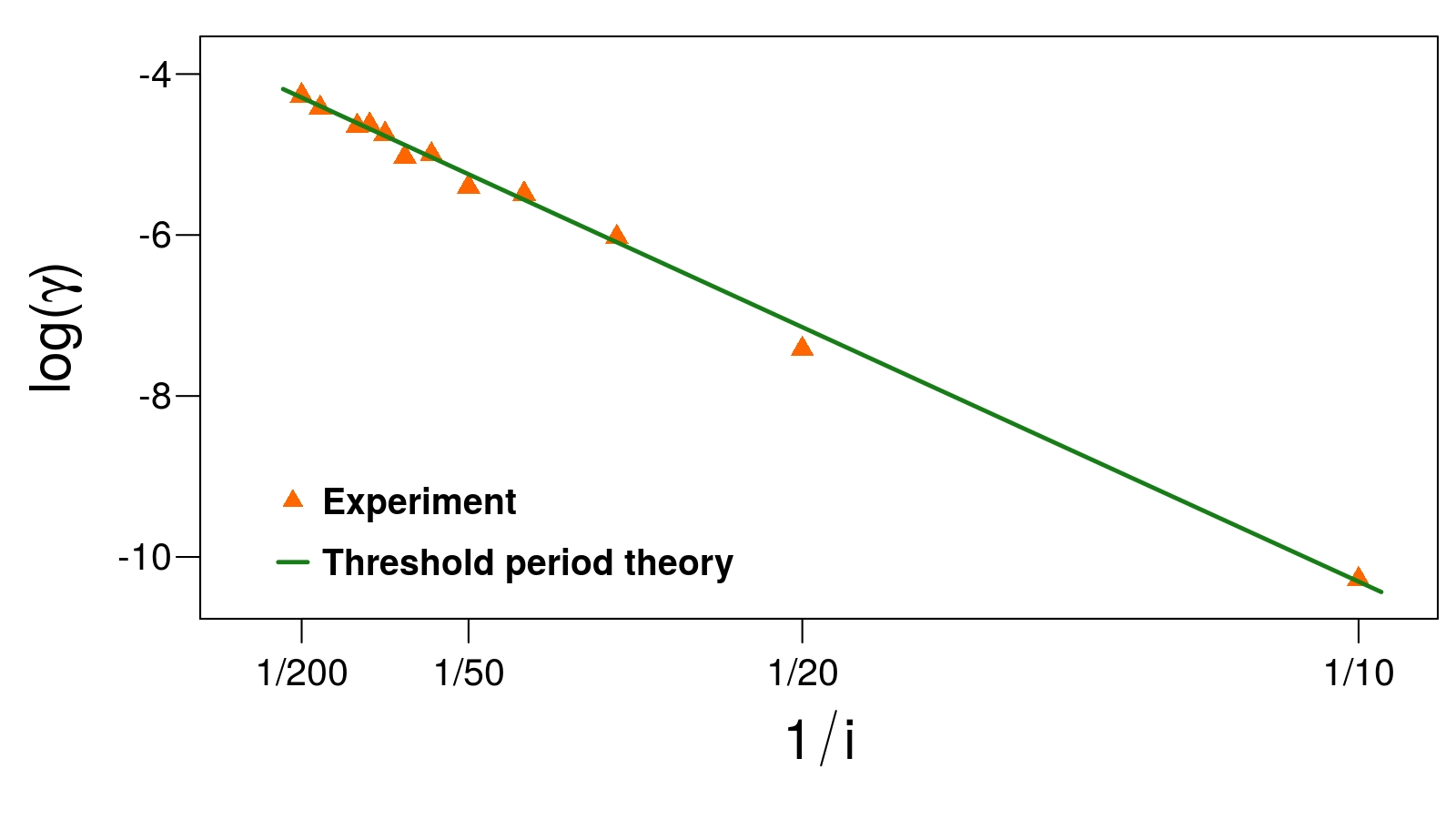}
\caption{
\textbf{Linear relationship between switching rate and inducer concentration on a logarithmic/inverse plot.} We replot data from Fig.~\ref{fig:transitionrate} in the form suggested by equation~(\ref{eq:tau_simp_rate}). The logarithm of the 
switching rate turns out to be in good approximation a linear function of the inverse external inducer concentration. The slope $-k_b \beta i_0$ and the 
intercept $\ln(d_t)-k_b \beta$ of this plot allow to directly read off the parameters $i_0$ and $k_b \beta$ from the 
experimentally measured switching rates (given the unbinding rate $d_t$, see SI Section \textit{2}).
} 
\label{fig:st_line}
\end{center} 
\end{figure}

{Typically, the repressor-DNA loop opens and closes multiple times before the \lac~regulatory region by chance remains free from repressors long enough to cause a switch in the phenotypic state. How frequently the regulatory region remains unbound for a period $\tau$} depends on two factors, the rate $k_b$ at which an operator site is located by and binds to a repressor and the rate $d_t$ at which a repressor completely unbinds from the promoter.  Assuming that a phenotypic switch happens when the \lac-regulatory region is unbound for a period exceeding $\tau$, the switching rate $\gamma$ can be calculated giving 
\begin{equation}
\label{eq:rate_switchon}
\frac{1}{\gamma} = e^{k_b \tau} \left[\left(\frac{1}{k_b}+\frac{1}{d_t}\right) -\frac{\tau}{e^{k_b \tau}-1}\right] \ ,
\end{equation}
see SI Sections \textit{2} and \textit{4} for details.
The switching rate $\gamma$ is high when $\tau$ is small (high external concentration of inducers) and when the rate of repressors binding to the operator $k_b$ is small. Thus, the threshold period theory explains the observation in theoretical models that speeding up the rate of repressor-operator binding and unbinding leads to a decrease of the switching rate~\cite{MorelliTanaseAllenWolde2008,keplerandelston,jaruref2}.
Taking the values of all other parameters from the literature, we use the maximum rate of inducer import per LacY molecule $\alpha_{tr} $ as the sole free parameter of our model, giving $\alpha_{tr}\approx 9.44/min$ 
(see SI Section \textit{4} for details). Fig.  {\ref{fig:transitionrate}} shows excellent agreement between the experimentally determined switching rate, the mechanistic model, and the theoretical switching rate curve Eq.  (\ref{eq:rate_switchon}).
If the gene is in the repressed state for most of the time ($k_b \gg d_t$), the switching rate (\ref{eq:rate_switchon}) simplifies to
\begin{equation}
\gamma = d_t e^{-k_b\tau}= d_t e^{-k_b\beta\left(1+ \frac{i_0}{i}\right)} \ ,\label{eq:tau_simp_rate}
\end{equation}
which is the rate of repressor unbinding $d_t$ divided by the expected number of unbinding-rebinding events $e^{k_b\tau}$ per switch (see SI Section \textit{4}). This expression predicts a linear relationship between 
the logarithm of the switching rate and the inverse external inducer concentration, see Fig.~\ref{fig:st_line}. 

\begin{figure*}
\begin{center}
\includegraphics*[width=.95\textwidth]{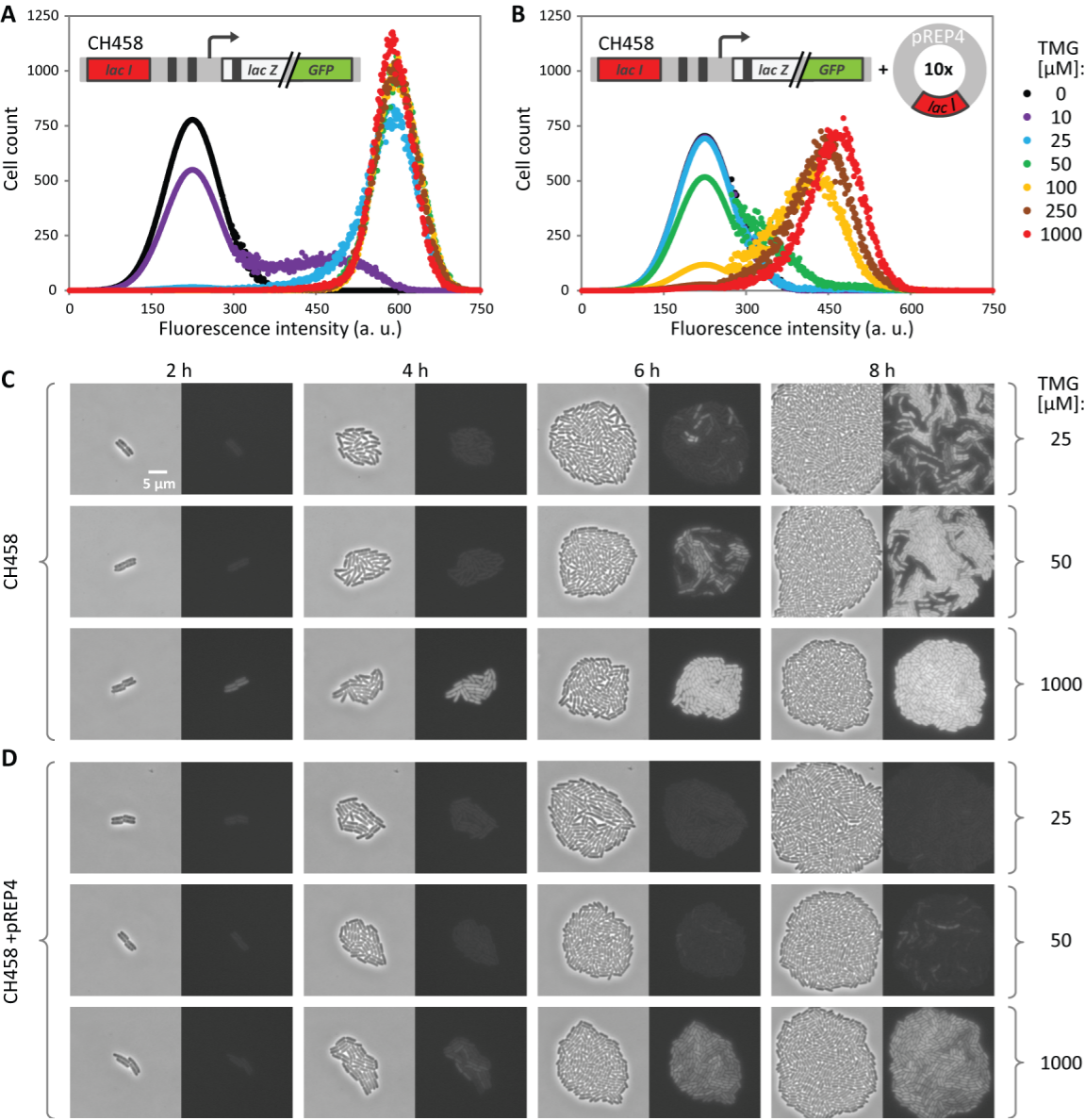}
\caption{{\bf The concentration of \lac-repressors affects the switching rate to the induced state.} 
{\bf (A-B)} Fluorescence distribution of a population of initially uninduced cells after 16h of incubation in the presence of a series of TMG concentrations (0, 10, 25, 50, 100, 250, 1000 $\mu$M TMG in black, purple, blue, green, yellow, brown, red, respectively). High fluorescence values indicate induced cells (colored dots) while low fluorescence values indicate uninduced cells (colored lines);
measurements in the low fluorescence range were fitted to a Gaussian distribution to include counts below the instrumental resolution limit (see also SI). 
Two strains were examined which differ in copy number of the \textit{lacI} gene. \textbf{(A)} In the original CH458 strain with a single copy of the \textit{lacI} gene, over the course of 16h virtually all cells switched to the induced state at TMG concentrations of 25 $\mu$M and above. \textbf{(B)} For the CH458+pREP4 strain containing 10 extra copies of the \textit{lacI} gene, no switching was observed at TMG concentrations of 25 $\mu$M and below. 
Intermediate TMG levels of $50$ $\mu M$ and $100$ $\mu M$ lead to transition of a fraction of the population
and only high TMG levels of 250 $\mu$M and above resulted in all cells switching to the induced state over 16 h. Note that an increased number of repressor molecules leads to reduced fluorescence intensity in the induced state, most likely due to short-term binding of repressor to the \lac-operator ~\cite{riggs1970,friedman1977}. \textbf{(C)} Time-lapse fluorescence microscopy shows 
individual cells of the original CH458 strain (left: phase-contrast image, right: fluorescence image). Cells were spotted on a semi-solid surface containing inducer concentrations of 25, 50 and 1000 $\mu$M TMG and then grew out to microcolonies. Still images from time-lapse movies show that the switching rate to the induced state increases with 
inducer concentration. Scale bar: 5 $\mu$m. \textbf{(D)} At 25 and 50 $\mu$M TMG, cells of the CH458+pREP4 strain with elevated LacI levels show markedly fewer transitions to the induced state compared to the original strain CH458.
\label{fig:plasmid}
}
\end{center}
\end{figure*}

If the phenotypic switch depends on the time period between repressor unbinding and rebinding, one expects the switching rate to depend on the rate at which repressors bind to the \lac-operator. Specifically, increasing number of \lac-repressors reduces the probability for the \lac~regulatory region to remain unbound for a period $\tau$, which should decrease the switching rate. We test this prediction of our theory by constructing an \textit{E. coli} strain with extra copies of the \textit{lacI} gene. The CH458+pREP4 strain expresses LacI from its original promoter on a plasmid containing a p15A origin of replication (10-12 copies per cell). We find the additional copies of the repressor strongly affect the switching behavior, leading to decreased switching seen both by flow cytometry (Fig. \ref{fig:plasmid}A-B) and by time-lapse fluorescence microscopy (Fig. \ref{fig:plasmid}C-D). This result cannot be explained by a scenario where the transition to the induced state is due to repressor-dissociation alone \cite{xieonmod}, since the dissociation event does not depend on the number of repressors present. 
Since in the uninduced state the \lac-operon is bound by a repressor most of the time, the 
most direct effect of an increased number of repressor molecules is to shorten the time repressor unbinding and repressor rebinding, in line with the threshold-period scenario above. 

\section*{Discussion}

Fluctuations taking a multistable gene regulatory system from one phenotypic state to another are among the most striking manifestations of stochasticity in biological systems. Here, we have shown that the switching rate curve, the rate of phenotypic switching as a function of a control parameter, is a powerful tool to identify the rate-limiting fluctuations that drive a particular transition and to
estimate molecular parameters associated with these fluctuations.


On the basis of precise measurements of the switching rate from the uninduced to the induced state of the lactose-uptake pathway in \textit{E. coli}, we used a detailed computational model of the \lac-pathway to formulate a theory of the rate-limiting fluctuation triggering \lac-induction. It is a combination of two chance events; (i) the repressor unbinding completely from its binding sites on the \lac-operon and (ii) the operator sites of the \lac-system staying free of repressors for a time sufficient to produce enough importers for positive feedback to kick in and take the cell into the induced state. We then altered the time period until \lac-repressor rebinding
by changing the concentration of LacI proteins and found that indeed shorter periods until \lac-repressor rebinding lead to decreased switching to the induced state.

Our result differs from the early picture of Novick and Weiner~\cite{novic}, where the production of a single molecule of permease is responsible for the phenotypic transition, or from the idea that bursts of \textit{lacY} translation are responsible. Our result complements and extends the recent result by Choi~\textit{et al.}~\cite{xieonmod} who, in single-molecule experiments, observed the repressor unbind completely from DNA prior to every
phenotypic transition. Hence, repressor dissociation is a \textit{necessary condition} for the phenotypic transition. We show that, however, repressor dissociation alone is not a
\textit{sufficient condition}; rather the operator needs to stay free of repressors for a threshold time period sufficient to import a critical amount of inducer into the cell. The second chance event -- all repressors \textit{not} binding to their binding sites for a specific period -- involves several molecules, and is not a single molecular event. This qualitative understanding of the phenotypic transition in the \lac-system has a large quantitative impact on the switching rate. Depending on the
external inducer concentration, the fraction of dissociation events leading to a phenotypic transition can be very small: at low inducer concentration the \lac~repressor completely dissociates from DNA more than thousand times before a single phenotypic transition occurs, see Fig. {\ref{fig:transitionrate}} and SI Section \textit{4}. Correspondingly, the rate of phenotypic switching and the rate of complete repressor unbinding differ by more than three orders of magnitude. 

%

Different multistable gene regulatory systems, from the~\textit{sonic hedgehog} pathway driving vertebrate organogenesis to competence development facilitating genetic transformation in \textit{B. subtilis}, share features such as feedback and hysteresis. Stochastic transitions between phenotypic states, on the other
hand depend on specific details which differ between systems. Understanding of the nature of the rate-limiting step of a phenotypic transition is a prerequisite for the design of synthetic switches with specific switching rates.


\section*{Competing interests}
The authors declare that they have no competing interests.

\section*{Acknowledgements}

We thank  A. Gordon for \textit{E. coli} strain CH458. We thank M. Leisner, P. Choi and L. Mirny for discussions and M. Markus for help with Fig. 1. 

\section*{Funding}
This work was supported by SysMO2 Noisy Strep.  Work in the Veening lab is also supported by the EMBO Young Investigator Program, a VIDI fellowship (864.12.001) from the Netherlands Organisation for Scientific Research, Earth and Life Sciences (NWO-ALW) and ERC starting grant 337399-PneumoCell.

\bibliography{lacbib}
\bibliographystyle{unsrt}

\end{document}